\documentclass[final,1p,times]{elsarticle}
\usepackage{amssymb}
\journal{Nuclear Physics A}

\begin{document}

\begin{frontmatter}



\title{p$K^+\Lambda$ final state: towards the extraction of  the $ppK^-$ contribution}
\author[8]{L.~Fabbietti}
\author[6]{G.~Agakishiev}
\author[7]{C.~Behnke}
\author[16]{D.~Belver}
\author[6]{A.~Belyaev}
\author[8]{J.C.~Berger-Chen}
\author[1]{A.~Blanco}
\author[7]{C.~~Blume}
\author[9]{M.~B\"{o}hmer}
\author[16]{P.~Cabanelas}
\author[6]{S.~Chernenko}
\author[10]{C.~~Dritsa}
\author[2]{A.~Dybczak}
\author[8]{E.~Epple}
\author[6]{O.~Fateev}
\author[1,a]{P.~Fonte}
\author[9]{J.~Friese}
\author[7]{I.~Fr\"{o}hlich}
\author[4,b]{T.~Galatyuk}
\author[16]{J.~A.~Garz\'{o}n}
\author[7]{K.~Gill}
\author[11]{M.~Golubeva}
\author[4]{D.~Gonz\'{a}lez-D\'{\i}az}
\author[11]{F.~Guber}
\author[4]{M.~Gumberidze}
\author[4]{S.~Harabasz}
\author[14]{T.~Hennino}
\author[10]{C.~~H\"{o}hne}
\author[3]{R.~Holzmann}
\author[9]{P.~Huck}
\author[6]{A.~Ierusalimov}
\author[11]{A.~Ivashkin}
\author[9]{M.~Jurkovic}
\author[5,c]{B.~K\"{a}mpfer}
\author[11]{T.~Karavicheva}
\author[3]{I.~Koenig}
\author[3]{W.~Koenig}
\author[3]{B.~W.~Kolb}
\author[2]{G.~Korcyl}
\author[16]{G.~Kornakov}
\author[5]{R.~Kotte}
\author[15]{A.~Kr\'{a}sa}
\author[7]{E.~Krebs}
\author[15]{F.~Krizek}
\author[2,14]{H.~Kuc}
\author[15]{A.~Kugler}
\author[11]{A.~Kurepin}
\author[6]{A.~Kurilkin}
\author[6]{P.~Kurilkin}
\author[6]{V.~Ladygin}
\author[8]{R.~Lalik}
\author[3]{S.~Lang}
\author[8]{K.~Lapidus}
\author[12]{A.~Lebedev}
\author[1]{L.~Lopes}
\author[7]{M.~Lorenz}
\author[9]{L.~Maier}
\author[1]{A.~Mangiarotti}
\author[7]{J.~Markert}
\author[10]{V.~Metag}
\author[7]{J.~Michel}
\author[7]{C.~M\"{u}ntz}
\author[8]{R.~M\"{u}nzer}
\author[5]{L.~Naumann}
\author[2]{M.~Palka}
\author[13,d]{Y.~Parpottas}
\author[3]{V.~Pechenov}
\author[7]{O.~Pechenova}
\author[3]{J.~Pietraszko}
\author[2]{W.~Przygoda}
\author[14]{B.~Ramstein}
\author[7]{L.~~Rehnisch}
\author[11]{A.~Reshetin}
\author[7]{A.~Rustamov}
\author[11]{A.~Sadovsky}
\author[2]{P.~Salabura}
\author[7]{T.~Scheib}
\author[7]{H.~Schuldes}
\author[8]{J.~Siebenson}
\author[15]{Yu.G.~Sobolev}
\author[8]{A.~Solaguren-Beascoa Negre}
\author[e]{S.~Spataro}
\author[7]{H.~Str\"{o}bele}
\author[7,3]{J.~Stroth}
\author[2]{P~Strzempek}
\author[3]{C.~Sturm}
\author[15]{O.~Svoboda}
\author[7]{A.~Tarantola}
\author[7]{K.~Teilab}
\author[15]{P.~Tlusty}
\author[3]{M.~Traxler}
\author[13]{H.~Tsertos}
\author[6]{T.~~Vasiliev}
\author[15]{V.~Wagner}
\author[9]{M.~Weber}
\author[5,c]{C.~Wendisch}
\author[5]{J.~W\"{u}stenfeld}
\author[3]{S.~Yurevich}
\author[6]{Y.~Zanevsky} 

\address[1]{LIP-Laborat\'{o}rio de Instrumenta\c{c}\~{a}o e F\'{\i}sica Experimental de Part\'{\i}culas , 3004-516~Coimbra, Portugal}
\address[2]{Smoluchowski Institute of Physics, Jagiellonian University of Cracow, 30-059~Krak\'{o}w, Poland}
\address[3]{GSI Helmholtzzentrum f\"{u}r Schwerionenforschung GmbH, 64291~Darmstadt, Germany}
\address[4]{Technische Universit\"{a}t Darmstadt, 64289~Darmstadt, Germany}
\address[5]{Institut f\"{u}r Strahlenphysik, Helmholtz-Zentrum Dresden-Rossendorf, 01314~Dresden, Germany}
\address[6]{Joint Institute of Nuclear Research, 141980~Dubna, Russia}
\address[7]{Institut f\"{u}r Kernphysik, Goethe-Universit\"{a}t, 60438 ~Frankfurt, Germany}
\address[8]{Excellence Cluster 'Origin and Structure of the Universe' , 85748~Garching, Germany}
\address[9]{Physik Department E12, Technische Universit\"{a}t M\"{u}nchen, 85748~Garching, Germany}
\address[10]{II.Physikalisches Institut, Justus Liebig Universit\"{a}t Giessen, 35392~Giessen, Germany}
\address[11]{Institute for Nuclear Research, Russian Academy of Science, 117312~Moscow, Russia}
\address[12]{Institute of Theoretical and Experimental Physics, 117218~Moscow, Russia}
\address[13]{Department of Physics, University of Cyprus, 1678~Nicosia, Cyprus}
\address[14]{Institut de Physique Nucl\'{e}aire (UMR 8608), CNRS/IN2P3 - Universit\'{e} Paris Sud, F-91406~Orsay Cedex, France}
\address[15]{Nuclear Physics Institute, Academy of Sciences of Czech Republic, 25068~Rez, Czech Republic}
\address[16]{LabCAF. F. F\'{\i}sica, Univ. de Santiago de Compostela, 15706~Santiago de Compostela, Spain} 
\address[a]{Also at ISEC Coimbra, ~Coimbra, Portugal}
\address[b]{Also at ExtreMe Matter Institute EMMI, 64291~Darmstadt, Germany}
\address[c]{Also at Technische Universit\"{a}t Dresden, 01062~Dresden, Germany}
\address[d]{Also at Frederick University, 1036~Nicosia, Cyprus}
\address[e]{Also at Dipartimento di Fisica Generale and INFN, Universit\`{a} di Torino, 10125~Torino, Italy}

%\author{L. Fabbietti$^{1}$ for the HADES Collaboration}

%\address{1: Excellence Cluster Universe, Technisch	e Universit\"{a}t M\"{u}nchen, Boltzmannstr. 2, D-85748, Garching, Germany}

\begin{abstract}
The reaction $p(@3.5\,GeV)+p\rightarrow p+\Lambda + K^+$ can be studied to search for the existence of kaonic bound states like $ppK^-$ leading to this final state. This effort has been motivated by the assumption that in p+p collisions the $\Lambda(1405)$ resonance can act as a doorway to the formation of the kaonic bound states.
The status of this analysis within the HADES collaboration, with particular emphasis on the comparison to simulations, is shown in this work and the deviation method utilized by the DISTO collaboration in a similar analysis is discussed. The outcome suggests the employment of a partial wave analysis to disentangle the different contributions to the measured $\mathrm{pK^+\Lambda}$ final state.
\end{abstract}

\begin{keyword}
$\Lambda(1405)$, kaonic bound state, meson-baryon interaction, partial wave analysis

\PACS
13.75,14.20

\end{keyword}

\end{frontmatter}


\section{Introduction}
\label{intro}
The study of the Kaon-nucleon interaction has triggered several experiments and theoretical calculations in the last two decades. From an experimental point of view, the kaon production has been investigated at intermediate energies ($\mathrm{E_{kin}= 1-4GeV}$) for heavy ion collisions and elementary reactions. Normally, the measured kinematic variables can be compared to transport models to infer information about the kaon-nucleus interaction. In this context, the $\Lambda (1405)$ resonance plays an important role. Indeed this baryon is theoretically described as a molecular state composed of either a $\mathrm{\bar{K}-p}$ or $\pi-\Sigma$ combination. Moreover, one expects that the production process and also the properties of the $\Lambda (1405)$ might differ upon the entrance reaction channel.  If we consider that the $\Lambda(1405)$ is partially composed by a $\mathrm{K^--p}$  bound state, by adding a additional proton we might obtain a $\mathrm{ppK^-}$ cluster \cite{Aka02}. This hypothesis also relies upon the fact that the kaon nucleon interaction is thought  to be strongly attractive \cite{Fuchs}. One could really think that the $\Lambda(1405)$ produced together with an additional proton might stick to it and form a $ppK^-$. Experimentally, we have addressed this issue by studying on the one hand the reaction $\mathrm{p+p\rightarrow \Lambda(1405) +K^+ +p}$ and on the other hand $\mathrm{p+p\rightarrow ppK^- +K^+\rightarrow p+\Lambda + K^+}$. In this work, we discuss the status of the analysis of the $\mathrm{p}\,\mathrm{K}^+ \Lambda$ final state. 

Our recent results about the $\Lambda(1405)$ production \cite{L1405Hades} show that the position of the maximum of the spectral function is found to be below $1390\,\mathrm{ MeV/c^2}$, suggesting a shift of the $\Lambda(1405)$ towards smaller masses with respect to the nominal value reported in the PDG. 
The analysis presented in \cite{L1405Hades} does not include the contribution of interferences between the $\Lambda(1405)$ and the I=0 phase space background, which could account for the shift and also modify the obtained differential cross-sections. 
Nevertheless, by neglecting interferences the angular distribution in the center of mass system (CMS) extracted for the $\Lambda(1405)$ indicates a rather isotropic production of the resonance, which is in agreement with the hypothesis of a rather large momentum exchange and a rather central p+p collision linked to this final state \cite{Has13}.
 
According to the theoretical predictions by \cite{Aka02}, the formation of the most fundamental of the kaonic  bound states ($ \mathrm{ppK^-}$) can happen in p+p collisions through the $\Lambda(1405)$ doorway. The underlying idea is that the $\Lambda(1405)$ being already a $\mathrm{K^-p}$ bound state, if this resonance is produced together with another proton and the relative momentum between the two particles is relatively small, the high attractive $K^-$-nucleon interaction might lead to the capture of a second proton by the $\Lambda(1405)$ and hence to the formation of a $\mathrm{ppK^-}$ molecule.
This scenario is predicted to be favored for p+p collisions at kinetic energies between $3-4\,\mathrm{GeV}$, where a large momentum transfer from the projectile to the target characterizes the dynamics and creates the optimal conditions for the formation of the kaonic cluster \cite{Aka02}.
From a theoretical point of view, the situation is rather controversial \cite{ppKTheo}. As summarized in \cite{Gal10}, different theoretical approaches predict the existence of a bound state like a $\mathrm{ppK^-}$, but the range of the predicted binding energies and width is rather broad and vary from $16-95\,\mathrm{MeV/c^2}$ and $34-110\,\mathrm{MeV/c^2}$ respectively. 
From an experimental point of view, signatures connected to the $\mathrm{ppK^-}$ have been collected by \cite{DIS1,Fin05}.
The result by the FINUDA collaboration \cite{Fin05} refers to measurement of stopped kaons on several solid targets and reports about a $ \mathrm{ppK^-}$ state with a binding energy of $115^{+6+3}_{-5-4}\,\mathrm{MeV}$ and a width of $67^{+14+2}_{-11-3}\,\mathrm{MeV}$; while the DISTO collaboration measured p+p reactions at $2.85\,GeV$ kinetic energy and found 
evidence for an exotic state with a binding energy of about $100\,\mathrm{MeV}$ and a width of $118 \pm 8\,\mathrm{MeV}$. 

Following the same assumptions discussed in \cite{DIS1}, we have carried out an analysis of the final state:
\begin{equation}
\label{pkl}
p+p  \rightarrow p+ K^+ +\Lambda  \rightarrow  p+K^+ + p + \pi^- 
\end{equation}
to investigate the possibility of having an intermediate state $\mathrm{p+p  \rightarrow ppK^- + K^+ }$ and the successive decay $\mathrm{ppK^-\rightarrow p+ \Lambda }$.

\section{Events Selection and Analysis}
The experiment was performed with the
{\bf H}igh {\bf A}cceptance {\bf D}i-{\bf E}lectron {\bf S}pectrometer (HADES) \cite{hades_nim}
at the heavy-ion synchrotron SIS18 at GSI Helmholtzzentrum f\"ur Schwerionenforschung in Darmstadt, Germany. 
A proton beam of $\sim 10^7$ particles/s with $3.5\, \mathrm{GeV}$ kinetic energy was incident on a liquid hydrogen target of $50\, \mathrm{ mm}$ thickness corresponding to $0.7\,\%$ interaction length.  
The data readout was started by a first-level trigger (LVL1) requiring a charged-particle multiplicity, $\mathrm{MUL}\,>3$, in the META system.
A total of $1.14\times10^9$ events was recorded under these experimental conditions.
The first analysis step consists of selecting events containing four charged particles ($p$, $\pi^-$, p, K$^+$).
Particle identification is performed employing the energy loss ($dE/dx$) of protons and pions  in the MDCs. The selection of the $\Lambda$ hyperon is carried out by exploiting the invariant mass of the $\mathrm{p- \pi^-}$ pairs and the cuts described in \cite{Sig1385}. A kinematic refit of the events containing a $\Lambda$ candidate, a proton and a third positive particle is first carried out, employing the energy and momentum conservation and also requiring the $\Lambda$ nominal mass for the selected $\mathrm{p}-\pi^-$ combination as constraints.
\begin{figure}[htb]
\label{kaon}
\begin{minipage}[t]{70mm}
      \begin{picture}(200,160)(0,0)
      \put(0,-2.6){ \includegraphics[width=0.88\textwidth,height=.7\textwidth]
          {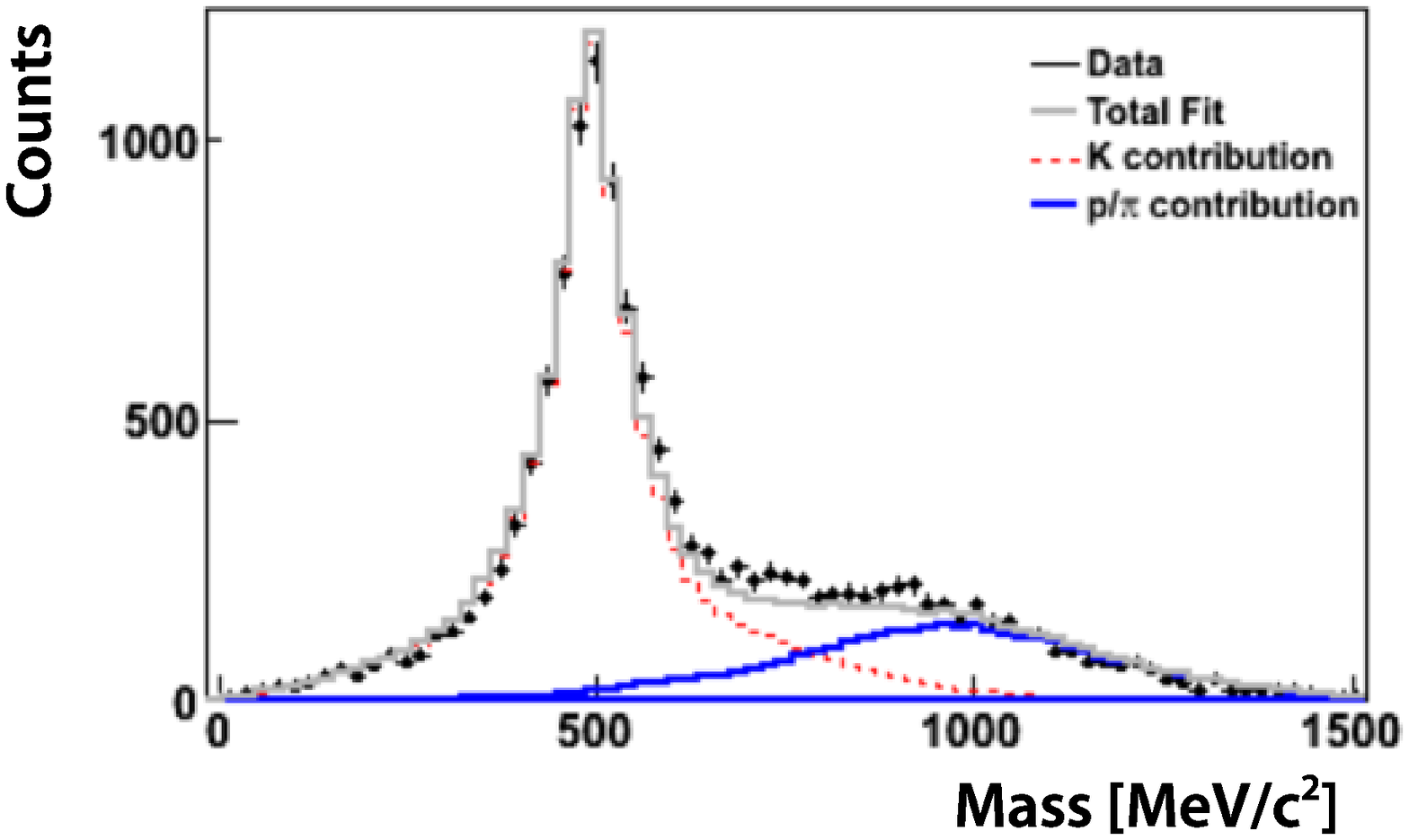}}
      \end{picture}
\end{minipage}
\hspace{\fill}
\begin{minipage}[t]{70mm}
      \begin{picture}(200,160)(0,0)
      \put(0,-2.6){ \includegraphics[width=.88\textwidth,height=.7\textwidth]
          {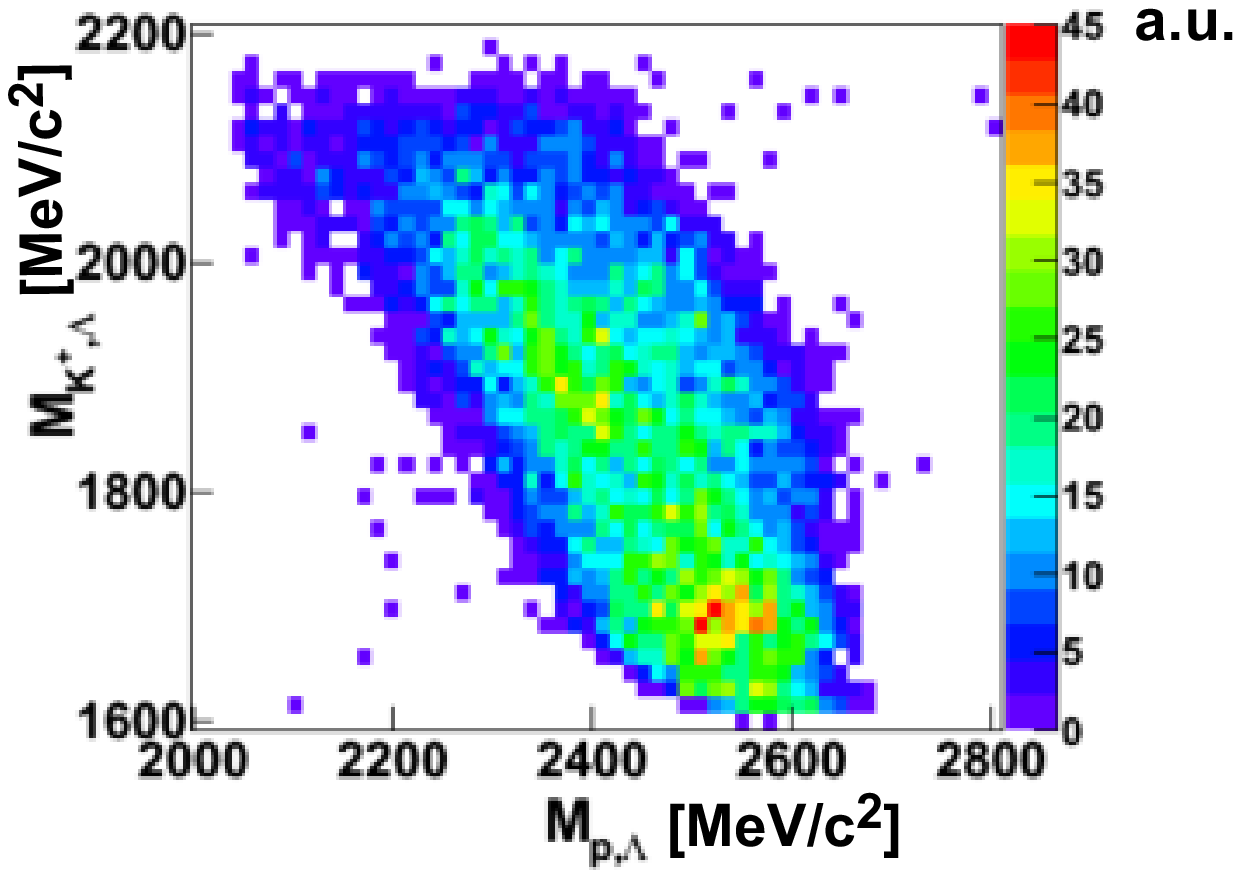}}
      \end{picture}
\end{minipage}
\caption{(Left)  Color online. Reconstructed mass of the kaon candidates via the measurement of the $\beta$ versus momentum. The full circles represent the experimental data, the red dashed line the contribution by the $K^+$ and the blue solid line the contribution from the protons. The gray solid line shows the global fit to the experimental data (see text for details). (Right) Color online. Correlation plot of the $\mathrm{K^+-\Lambda}$ ($\mathrm{M(K^+-\Lambda)}$) invariant mass as a function of the $\mathrm{p-\Lambda}$ invariant mass the experimental data for the exclusive reaction $p+p\rightarrow p+ K^+ +\Lambda$. }
\end{figure}

The kinematic refit allows to select events corresponding to the $\mathrm{p+K^+ +\Lambda}$ final state. A total statistic of $11.000$ events is extracted and the mass of the third positive particle is shown in Fig.~1 (left panel). The full circles represent the experimental data corresponding to the selected $\mathrm{p+K^++\Lambda}$ events after the kinematic refit, the red dashed and the blue solid line correspond to full-scale simulations  and represent the response to the kaon and proton signal respectively. The simulation are not absolutely normalized but the scaling factor is chosen such to reproduce to experimental distribution.
One can see that the exclusive analysis allows a good $\mathrm{K^+}$ identification with a rather low contamination by protons, which translates into a signal to background ratio of about $15$. Within a $3\,\sigma$ cut around the nominal $\mathrm{K^+}$ mass, a background contribution of about $2\,\%$ has been estimated.
Fig.~1 (right panel) shows a correlation plot for the selected reaction $\mathrm{p+p \rightarrow p+\Lambda +K^+}$ where the $\mathrm{K^+-\Lambda}$ ($\mathrm{M(K^+-\Lambda)}$) invariant mass is shown as a function of the $\mathrm{p-\Lambda}$ invariant mass ($\mathrm{M(p-\Lambda)}$) within the HADES acceptance and before the efficiency corrections. This distribution gives an impression of the phase space coverage which is accessible for this final state using the HADES spectrometer.
The analysis method discussed in \cite{DIS1} relies upon the method of the deviation plot. The experimental $\mathrm{pK^+\Lambda}$ Dalitz plot is divided by the Dalitz plot obtained by simulating the production of the $\mathrm{pK^+\Lambda}$ final state by pure phase space emission. The projection of the so obtained ratio along the $\mathrm{M(p-\Lambda)^2}$ shows a large bump, and this bump is interpreted in \cite{DIS1} as the evidence of an exotic state. By fitting the deviation plot obtained for $\mathrm{M(p-\Lambda)}$ and the $\mathrm{K^+}$ missing mass ($\mathrm{MM(K^+)}$) with a Gaussian superimposed to a linear background,  a structure with the mass  $\mathrm{M_X=\,2.265 \pm 0.002\,GeV/c^2}$ and a width $\mathrm{\Gamma_X =\, (0.118\pm\, 0.008)\, GeV/c^2}$ has been identified and associated to a bound state of two protons and a $\mathrm{K^-}$.

\begin{figure}[h]
\centering
\includegraphics*[width=13.2cm]{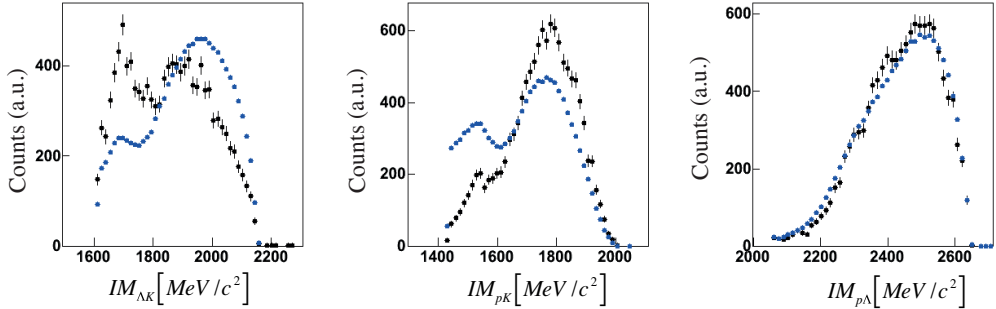}
\caption{Color online. The full circles in black show the experimental distribution for the invariant mass of the particle pairs: $\mathrm{\Lambda K}$ (a), $\mathrm{pK}$ (b) and $\mathrm{p\Lambda}$ (c). The full circles in blue show the same distributions obtained from the phase-space simulation of the $\mathrm{pK\Lambda}$ final state.}
\label{Sigma}
\end{figure}

It is clear that such a method does not take into account the role played by resonances like $\mathrm{N^*}$, the interferences among the different intermediate states and their contribution to the experimental spectrum.
As a first step, we would like to address the comparison of the experimental data to the $\mathrm{pK^+\Lambda}$ phase space simulation.
We have carried out full scale simulation of the $\mathrm{pK^+\Lambda}$ final state by pure phase space emission within the HADES acceptance and we have compared these simulations to the experimental data within the acceptance. Fig.~\ref{Sigma} shows the three invariant mass spectra of the $\mathrm{pK^+\Lambda}$ final state. The full circles in black show the experimental distributions for the invariant mass of the particle pairs: $\mathrm{M(\Lambda K^+)}$ (a), $\mathrm{M(pK^+)}$ (b) and $\mathrm{M(p\Lambda)}$ (c). The full circles in blue show the same distributions obtained from the phase-space simulation of the $\mathrm{pK^+\Lambda}$ final state. One can see that the invariant mass distributions differ evidently, especially the $\mathrm{\Lambda K}$  and $\mathrm{pK}$ invariant mass distribution. If we compare the phase space simulations and the experimental data on the base of the angular distribution in the CMS, Gottfried-Jackson and helicity reference frames defined analog to \cite{Sig1385}, the disagreement is visible as well.  
\begin{figure}[h]
\centering
\includegraphics*[width=14.2cm]{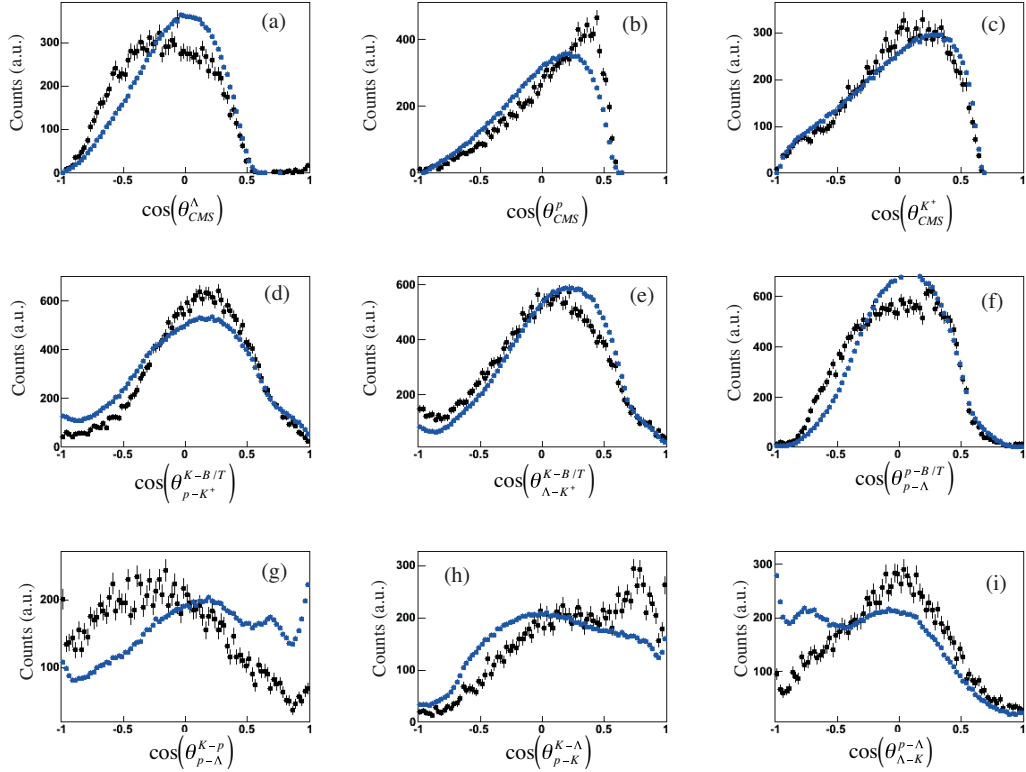}
\caption{Angular distributions for the production in CMS of $\Lambda$, $p$ and $K^+$ (top row: $a:\,\Theta^{\Lambda}_{CMS},\,b:\Theta^{p}_{CMS}, \,c:\,\Theta^{K^+}_{CMS}$), Gottfried-Jackson (middle row $d:\,\Theta^{K-B/T}_{p-K},\,e:\,\Theta^{K--B/T}_{\Lambda-K},\,f:\,\Theta^{p-B/T}_{p-\Lambda}$) and helicity  angles (bottom row: $g:\,\Theta^{K-p}_{p-\Lambda},\, h:\,\Theta^{K-\Lambda}_{p-K},\,i:\,\Theta^{p-\Lambda}_{K-\Lambda}$) angle frames. The full circles in black show the experimental data and those in blue the same distributions obtained from the phase-space simulation of the $\mathrm{pK\Lambda}$ final state.}
\label{angles}
\end{figure}
Fig.~\ref{angles} shows the angular distribution for the experimental data and the phase space simulations within the HADES acceptance for all the combinations in the CMS, Gottfried-Jackson and helicity reference frames. The fact that the phase space simulations do not show isotropic and symmetric distributions is partially due to the geometrical acceptance of the spectrometer for the studied reaction, but these effect are under control in the simulation package.
The same disagreement is found if the momentum distribution of the single particles are compared. 
These comparisons show that the deviation between the phase space distribution and the experimental $\mathrm{pK^+\Lambda}$ final states can not be explained by the incoherent sum of the phase space distribution with a single additional resonant state in the $\mathrm{p-\Lambda}$ channel. For this reason a deviation plot would be very difficult to interpret. 

\section{Contribution by the N$^*$ Resonances}
As suggested by the experimental invariant mass distribution of the $\mathrm{K^+-\Lambda}$ pairs and as visible in Fig.~\ref{Sigma}, the contribution by intermediate $\mathrm{N^*}$ resonances decaying into $\mathrm{K-\Lambda}$ pairs should be considered. The left panel of Fig.~\ref{Sigma} shows two broad peaks around $1700$ and $1900\,\mathrm{MeV/c^2}$ and suggests the presence of at least two N$^*$, but due to the acceptance effects this hypothesis needs to be verified via full-scale simulations.
\begin{figure}[h]
\centering
\includegraphics*[width=12.2cm]{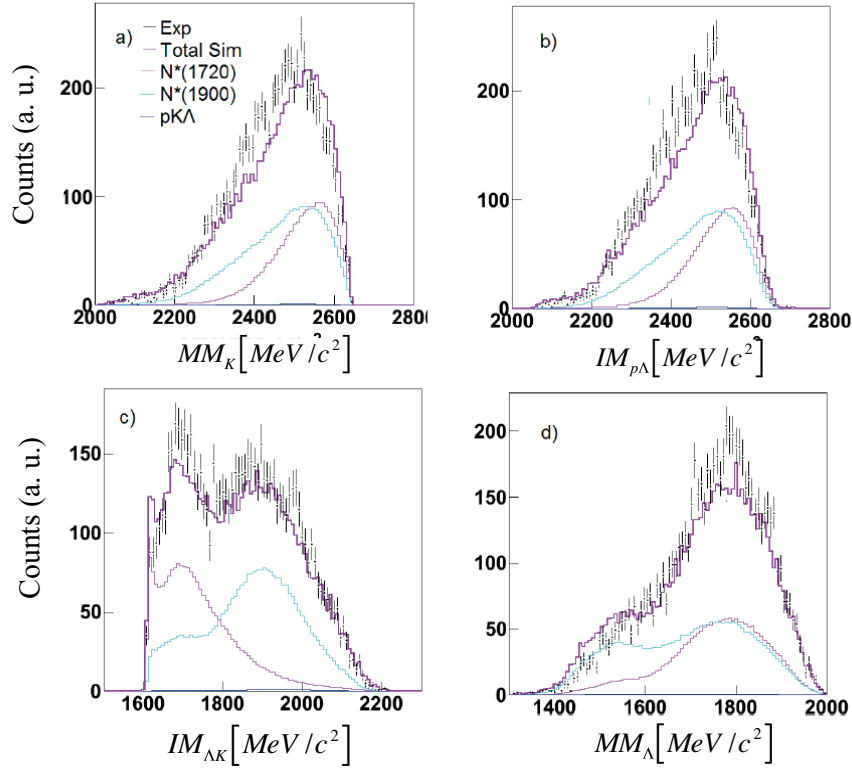}
\caption{Color online. K$^+$ Missing mass (a), $p-\Lambda$ invariant mass (b), $\Lambda-K$ invariant mass (c) and $\Lambda$ missing mass (d) distributions. The black dots show the experimental data, the cyan and the magenta histograms show the contributions by the $N^*(1900)$ and $N^*(1720)$ resonances obtained from full scale simulations, the violet histogram shows the total sum of the simulations.}
\label{NCock}
\end{figure}
As a first attempt, simulations have been carried out including the incoherent sum of four $\mathrm{N^*}$ resonances with a mass of $1650,\,1720,\,1900$ and $2190\,\mathrm{MeV/c^2}$ together with the phase-space production of the $\mathrm{pK^+\Lambda}$ state. The parameters of the resonances used in the simulations are summarized in Table 1. The choice of these resonances is rather arbitrary and constrained by the fact that exploiting a mere incoherent simulation model will not allow to distinguish the contributions by the $\mathrm{N^*(1710)}$ and $\mathrm{N^*(1720)}$ or other resonance pairs lying at higher masses with a mass difference lower than $\mathrm{20\,MeV/c^2}$, being all these states rather broad.
\begin{table}[t]
\begin{center}
\begin{tabular}{|c||c|c|c|c|}
\hline
$N^*$ Mass [$\mathrm{MeV/c^2}$] & 1650& 1720& 1900 &2190 \\
\hline
$N^*$ Width [$\mathrm{MeV/c^2}$] & 165 & 200 & 180 & 500\\
\hline
PDG Evidence &***&**& &* \\
\hline
\end{tabular}
\end{center}
\label{t1}
\caption{Masses and widths of the $N^*$ resonances employed in the simulations. The values are taken from the PDG \cite{PDG11}.}
\end{table}

The strength of the different contributions has been varied such to reproduce as good as possible the experimental data. Fig.~\ref{NCock} shows the final result, after the optimization of the simulation cocktail, the $\mathrm{K^+}$ missing mass  (a), $\mathrm{p-\Lambda}$ invariant mass (b), $\mathrm{\Lambda-K}$ invariant mass (c) and $\Lambda$ missing mass (d) distributions are displayed. The black dots represent the experimental points within the HADES acceptance, the cyan and magenta histograms the contributions from the $\mathrm{N^*(1900)}$ and $\mathrm{N^*(1720)}$ resonances respectively, while the violet histogram correspond to the total simulated distributions. The contributions from the other two $\mathrm{N^*}$ resonances included in the full-scale simulations is set to 0 by the minimization procedure and also the contribution by the pure phase space production amounts to only $1.5\%$  of the total yield and is not clearly visible in Fig.~\ref{NCock}. The contribution by the $\mathrm{N^*(1720)}$ and $\mathrm{N^*(1900)}$ resonances amounts to $41.5\%$ and $57\%$ respectively and a total $\chi^{2}$ value of $3.2$ is obtained by the comparison of the simulated distribution to the experimental data for the kinematic variables shown in Fig.~\ref{NCock}.
The $\mathrm{M(\Lambda-K^+)}$ distribution shows a much improved agreement between the simulations and the experimental data, if compared to the distributions discussed in Fig.~\ref{Sigma}, and the two structures can mainly be associated to the contribution of the $\mathrm{N^*(1720)}$ and $\mathrm{N^*(1900)}$ resonances. The $\mathrm{\Lambda}$ missing mass distribution shows a similar qualitative agreement between the simulation and the experimental data, in particular the presence of the $\mathrm{N^*(1900)}$ resonance seems mandatory to describe the low missing mass region.
On the other hand, the incoherent simulation employed here, that does not even contain the proper angular distribution of the different final states, does not aim a quantitative determination of the different $\mathrm{N^*}$ contributions. A more compete analysis in this direction is currently being carried out.
When looking at the $\mathrm{p-\Lambda}$ invariant mass (Fig.~\ref{NCock} (b)), the simulated distribution is shiftd to the right hand side of the mass range, probably due to the fact that the dynamic of the reaction is not completely described by simulations.
Indeed, one has to point out that the experimental angular distributions in the CMS, Gottfried-Jackson and helicity reference frames can not be described by the new simulations including the $\mathrm{N^*}$ resonances, implying that interferences among the different intermediate states might play an important role and should be accounted for but also because the simulations so far have not been weighted with the correct production and decay angular distributions. 
\begin{figure}[h]
\centering
\includegraphics*[width=13.2cm]{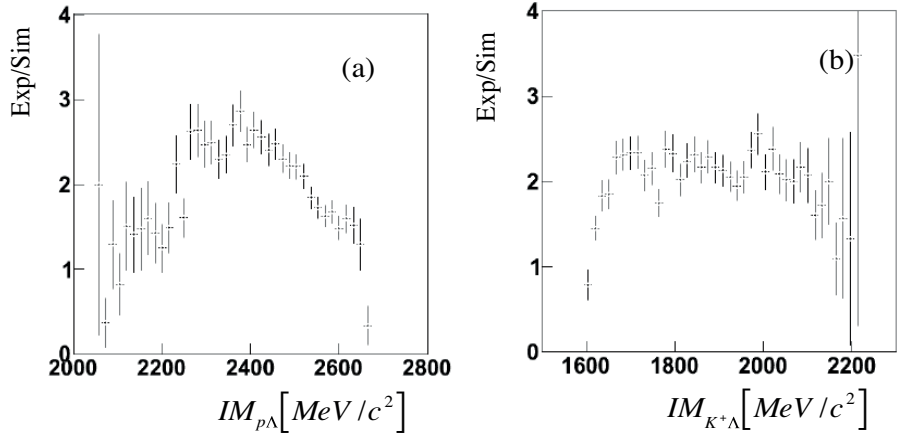}
\caption{Deviation plot for the $p-\Lambda$ (a) and $K-\Lambda$ (b) invariant mass distribution.}
\label{DevN}
\end{figure}
If we want to compare the experimental correlation plot shown in Fig.~1 (right panel) to the new simulations obtained adding incoherently the phase space production of the $pK^+\Lambda$ final state to the $\mathrm{N^*}$ contribution, we can build a deviation plot by dividing the experimental data with the simulation. The projections of this ratio on the $\mathrm{p-\Lambda}$ and $\mathrm{\Lambda-K^+}$ invariant mass axis is shown in Fig.~\ref{DevN}, (a) and (b) respectively. As one can see, the distribution for the $\mathrm{\Lambda-K^+}$ ratio is rather flat, while for the $\mathrm{p-\Lambda}$ invariant mass ratio the shift of the simulated distribution to the right hand side of the spectrum with respect to the experimental distribution, as visible in Fig.~\ref{NCock} (b), generates a broad bump in the deviation plot. Hence this bump can not be directly attributed to a resonance since the shift of the two spectra, which is also visible in the $\mathrm{K^+}$ missing mass distribution (Fig.~\ref{NCock} (a)), can be due to the fact that the simulation model that has been used for the comparison does not include basic features of the $\mathrm{pK^+\Lambda}$ production.

It has to be pointed out that several attempts have been made to model non isotropic angular distribution for the $\mathrm{N^*}$ resonances following the same line of reasoning as shown in the analysis in \cite{COSY_10}, but no solution was found which enables to reproduce the experimental data. New studies employing  partial wave analysis have been started and look very promising. A detailed modeling of the experimental data is also necessary to extract a valid acceptance correction, since the geometrical acceptance of the HADES spectrometer is not $100\%$.
The DISTO results assign the signature to the exotic state after a cut on the polar angle of the final state proton ($\mathrm{|{\cos \theta_{CMS}}| \geq 0.6}$) in order to suppress the phase space
 production contribution. This cut would not affect at all the HADES data, since small polar angles for final state protons are not accessible for this colliding system due to the limited 
 geometrical acceptance of the HADES spectrometer in the forward direction.  Moreover, our results stay the same even if a further cut on the
  $K^+$ emission angle ($\mathrm{-0.2 <  \cos \theta_{K^+} < 0.4} $), as employed in the DISTO analysis to improve the S/B ratio, is applied.
  
\section{Summary}
We have shown the analysis of the reaction $\mathrm{p+p\rightarrow p+\Lambda +K^+}$ for an incoming beam with a kinetic energy of $\mathrm{3.5\,\mathrm{GeV/c^2}}$ measured with the HADES spectrometer. A high purity sample of about $11.000$ exclusive $\mathrm{pK^+\Lambda}$ events has been extracted and the invariant mass correlation plot and the relative one dimensional projection have been 
compared to full scale simulation with a pure phase-space event generator.
The comparison shows that the phase-space simulations can not describe the experimental  missing mass, invariant mass and angular distributions. The disagreement can not be overcome by adding the contribution of one resonance in the $\mathrm{p-\Lambda}$ decay channel with a mass around $2300\,\mathrm{MeV/c^2}$. The $K^+-\Lambda$ invariant mass shows a clear contribution by at least two $\mathrm{N^*}$ resonances to the analyzed final state. A dedicated full-scale simulation, including additionally to the $\mathrm{pK^+\Lambda}$ phase-space distribution the contribution from $\mathrm{N^*(1720)}$ and $\mathrm{N^*(1900)}$ achieves a better description of the experimental data, but still fails to describe the angular distributions. The deviation plot in the $\mathrm{p-\Lambda}$ invariant mass distribution shows a wide bump around $2400\,\mathrm{MeV/c^2}$ that seems to be originated from a shift in the kinematic of the simulation respect to the experimental data. This observation jeopardizes the solidity of the deviation method exploited to extract the DISTO $\mathrm{ppK^-}$ signal.
Currently the partial wave analysis method is being investigated to include the interferences among the $\mathrm{N^*}$ resonances and all the other intermediate states contributing to the $pK^+\Lambda$ final state.

The HADES collaboration gratefully acknowledges the support by the grants LIP Coimbra, Coimbra (Portugal) PTDC/FIS/113339/2009, SIP JUC Cracow, Cracow (Poland): N N202 286038 28-JAN-2010 NN202198639 01-OCT-2010, FZ Dresden-Rossendorf (FZD), Dresden (Germany) BMBF 06DR9059D, TU M�nchen, Garching (Germany) MLL M\"unchen: DFG EClust 153, VH-NG-330 BMBF 06MT9156 TP5 GSI TMKrue 1012 NPI AS CR, Rez, Rez (Czech Republic) MSMT LC07050 GAASCR IAA100480803, USC - S. de Compostela, Santiago de Compostela (Spain) CPAN:CSD2007-00042, Goethe-University, Frankfurt (Germany): HA216/EMMI HIC for FAIR (LOEWE) BMBF:06FY9100I GSI F\&E.

\end{document}